\pdfoutput=1

\documentclass{article}

\usepackage{arxiv}

\usepackage{times}

\usepackage{multicol}
\usepackage[bookmarks=true]{hyperref}

\usepackage{graphicx}
\usepackage{amsfonts}         
\usepackage{amsmath} 
\usepackage{amssymb} 
\usepackage{amsthm}
\usepackage[noend]{algpseudocode}
\usepackage{algorithm}
\usepackage{diffcoeff}
\usepackage{subfigure}
\usepackage{tabularx}
\usepackage[symbol]{footmisc}
\usepackage{mathtools}
\usepackage{dsfont}
\usepackage{cases}
\usepackage{bigints}
\usepackage{cite}
\usepackage{url}
\usepackage{verbatim}
\usepackage{xcolor}

\DeclareMathOperator{\vecop}{vec}

\title{Desensitization and Deception in Differential Games with Asymmetric Information\thanks{This work has been submitted to a journal for possible publication. Copyright may be transferred without notice, after which this version may no longer be accessible.}}

\author{Vinodhini Comandur\thanks{Equal contributions}\\
Georgia Tech\\
\texttt{vinodhini@gatech.edu}\\
  \And
  Tulasi Ram Vechalapu\footnotemark[2]\\
  University of Alabama in Huntsville\\
  \texttt{tv0013@uah.edu}\\
  \And
  Venkata Ramana Makkapati\footnotemark[2]\\
  Honda Aircraft Company\\
  \texttt{vmakkapati3@gmail.com}\\
  \And
  Panagiotis Tsiotras\\
  Honda Aircraft Company\\
  \texttt{tsiotras@gatech.edu}\\
  \And
  Seth Hutchinson\\
  Georgia Tech\\
  \texttt{seth@gatech.edu}
}

\begin{document}

\maketitle

\begin{abstract}
Desensitization addresses safe optimal planning under parametric uncertainties by providing sensitivity function-based risk estimates.
This paper expands upon the existing work on desensitization in optimal control to address safe planning for a class of two-player differential games. 
In the proposed game, parametric uncertainties correspond to variations of the model parameters for each player about their nominal values. 
The two players in the proposed formulation are assumed to have perfect information about these nominal parameter values. 
However, it is assumed that only one of the players has complete knowledge of the actual parameter value, resulting in information asymmetry in the proposed game.
This lack of knowledge regarding the parameter variations is expected to result in state constraint violations for the player with an information disadvantage.
In this regard, a desensitized feedback strategy that provides safe trajectories is proposed for the player with incomplete information. 
The proposed feedback strategy is evaluated for instances involving a single pursuer and a single evader with an uncertain moving obstacle, where the pursuer is assumed to only know the nominal value of the obstacle's speed.
At the same time, the evader knows the obstacle's true speed, and also the fact that the pursuer knows only the nominal value of the obstacle's speed.
Subsequently, deceptive strategies are proposed for the evader, who has an information advantage, and these strategies are assessed against the pursuer's desensitized strategy.
\end{abstract}


\section{Introduction}

Games with asymmetric information involve instances where one player knows something the other player does not know \cite{cardaliaguet2012information}. 
These instances are particularly evident during military engagements, where the ``fog of war" plays a critical role in the decision-making process \cite{tryhorn2023modeling}.
In such engagements, two distinct behaviors can be observed \cite{cronin2008impenetrable}. 
The \emph{more informed player} tends to resort to deceptive strategies with the aim of imposing losses on its opponent. 
The \emph{less informed player} tends to mitigate its losses due to lack of information by choosing a risk-averse strategy.
This work models the two behaviors, risk aversion and deception, for a class of two-player differential games by making use of sensitivity functions.

From a game-theoretic standpoint, the term \emph{information} correlates to the structure of the game, including the game environment, the opponent's intentions (payoff functions), and the players' limitations (state and control constraints).
The aspect of information asymmetry among agents has been of great interest in the field of economics \cite{akerlof1970market, spence1978job, stiglitz1981credit}.
Cardaliaguet was one of the first to study differential games with asymmetric information and analyzed scenarios where players do not have the same information regarding a random terminal payoff function \cite{cardaliaguet2007differential}.
A complete analysis of the existence and characterization of a value function for such games can be found in \cite{buckdahn2011some}. 
Subsequently, Larsson et al. \cite{larsson2018nash} showed that for a given zero-sum game, information asymmetry always transforms the original game into a non-zero-sum game.
Using the concept of common information, Nayyar et al. characterized and computed the Nash equilibria for a class of asymmetric games by formulating an equivalent symmetric game \cite{nayyar2013common, gupta2014common}, and later applied the proposed approach to address security in cyber-physical systems \cite{gupta2016dynamic}.

In this work, information asymmetry corresponds to the lack of exact knowledge about the environment, captured via a set of uncertain model parameters in the case of one of the two agents.
Consequently, based on observations from real-world strategic engagements \cite{howard1990soft, hamilton2002challenges}, it is assumed that the agents would deviate from their original intentions, and resort to \textit{risk-averse} or \textit{deceptive} strategies by updating their respective payoff functions.
Each agent's choice of its own updated payoff function is not communicated to its adversary, which is another form of information asymmetry that is introduced in the proposed game.

In this paper, we propose risk estimates for the player having a lesser environmental awareness using the theory of desensitized optimal control (DOC) \cite{sarabu2020safe}.
DOC-based techniques employ sensitivity functions to estimate the expected first-order variations of a desired function under parametric uncertainties \cite{ramana2018doc,makkapati2020cdoc,makkapati2021games}.
Previously, these techniques were employed to reduce landing errors under uncertainties and perturbations for re-entry spacecraft \cite{shen2010desensitizing, hu2016desensitized, lou2016robust, makkapati2021desensi}. 
Along a similar vein,
Oyler et al. considered the sensitivity of the boundaries of dominance regions (safe-reachable sets) under variations in players' positions and obstacle positions in pursuit-evasion (PE) games \cite{oyler2015dominance, oyler2016contributions}.
The risk function proposed in \cite{oyler2016contributions} captures the probability that an opponent dominates a given point in Euclidean space.

The proposed approach is an extension of our earlier work on desensitization for optimal control problems \cite{sarabu2020safe} to the case of differential games.
Preliminary results of the proposed approach were discussed in  \cite{makkapati2022desensi} along with a proof-of-concept simulation involving a two-agent PE scenario with an uncertain dynamic obstacle.
In this paper, we introduce a novel way to incorporate deception in the aforementioned two-agent PE game with asymmetric information along with a sensitivity function-based mitigation strategy for the player with an information disadvantage.
This paper also provides comprehensive simulations that illustrate the limitations of desensitization and deception.

The impact of the information structure on the players' strategies in pursuit-evasion games has been a topic of interest for many years. 
Gurel-Gurevich proved that a value function exists for a class of discrete-time PE games with incomplete information \cite{gurel2009pursuit}.
A particular formulation where the evader observes just the initial conditions while the pursuer observes the instantaneous relative distance between the two opponents with additive noise was analyzed by Hexner et al. in~\cite{hexner2019}.
One of the most intuitive ways to address information uncertainty is to consider a state estimator.
Various heuristic pursuit strategies were evaluated for PE games with incomplete information, since obtaining optimal strategies for such instances is elusive \cite{antoniades2003}.
In this regard, estimation-based approaches for pursuit-evasion were considered in~\cite{mizukami1980state, cavalieri2014incomplete}, and were subsequently extended to orbital PE games in~\cite{shen2015pursuit, linville2020linear, ye2021multiple}.
Instances of lack of information were considered in visibility-based pursuit-evasion, and solutions to optimally track and capture an unpredictable evader were analyzed in~\cite{guibas1999visibility, gerkey2006visibility, bhattacharya2010existence}. 
Epistemic models for PE games with visibility constraints were analyzed by Huang et al in~\cite{Huang2011}.  
Information delays in pursuit-evasion were considered by Shinar and Glizer in~\cite{shinar1999solution}.
Finally, deception has been studied for territorial defense applications \cite{asgharnia2020deception, asgharnia2022learning}, where the invader hides information regarding its goals from the defenders .

The proposed PE game analyzes instances involving a pursuer, an evader, and a moving obstacle whose exact position and velocity are precisely known only to one of the players, namely, the evader.
The pursuer knows only the initial position of the obstacle and has some partial information about the obstacle's velocity.
On the other hand, the evader is aware of the pursuer's lack of information regarding the obstacle's whereabouts.
In contrast to the existing literature on PE games involving obstacles and information asymmetry \cite{nayyar2013common, gupta2014common, gupta2016dynamic, oyler2015dominance, oyler2016contributions}, the players in the proposed problem formulation react to the underlying information asymmetry by considering updated objectives that are not communicated to their respective opponents.
Since the evader knows the true value of the obstacle's velocity, she tries to drive the pursuer toward the obstacle using a deceptive strategy.
In order to capture the evader while avoiding the dynamic uncertain obstacle, the pursuer employs a desensitized strategy that provides safer trajectories to achieve his goal.
In this work, we consider a simple receding-horizon approach to obtain optimal solutions for the players in the proposed PE game.

The contributions of this paper are as follows: 
a) a novel problem formulation is introduced for differential games with asymmetric information by capturing the lack of environmental awareness using uncertain model parameters; 
b) a sensitivity function-based risk estimate is introduced for differential games to generate safer trajectories that reduce constraint violations for the less informed player; 
c) a deceptive strategy is proposed for the more informed player;
and, finally, 
d) the efficacy of the proposed strategies (both individually and combined) is verified for a class of pursuit-evasion games via simulations that showcase their applicability.
To the best of the authors' knowledge, the proposed formulation and the solution approach are novel to differential games of pursuit-evasion type.

The paper is organized as follows. 
Section \ref{sec:prelims} presents the proposed formulation for differential games with asymmetric information.
Section \ref{sec:decept} provides a mathematical framework for defining a deceptive agent in the realm of the proposed formulation.
Section \ref{sec:cdp} discusses constrained desensitized planning (CDP), and provides a risk metric to counter information disadvantage and deception in asymmetric differential games. 
Section \ref{sec:pe} analyzes deception and CDP using a pursuit-evasion example involving an uncertain moving obstacle. 
Section~\ref{sec:conclude} concludes the paper and provides some directions for future work.

\section{An Asymmetric Differential Game}
\label{sec:prelims}

Consider a two-player differential game with dynamical equations
\begin{align}
    &\dot{x}_p(t) = f_p(x_p(t),u(t),t), \qquad x_p(t_0) = x^0_{p}, \label{eq:p_dyn}\\
    &\dot{x}_e(t) = f_e(x_e(t),v(t),t),  \qquad x_e(t_0) = x^0_{e}, \label{eq:e_dyn}\\
    &\dot{x}_w(t) = f_{w}(x_w(t),\rho,t),  \qquad \quad x_w(t_0) = x^0_{w}, \label{eq:env_dyn}
\end{align}
where $t \in [t_0,\,t_f]$ denotes current time, $t_0$ is the initial time, $t_f$ is the time when the game ends,
$x_p(t) \in \mathbb{R}^{n_p}$ denotes the state vector of one of the two players  at time $t$, referred to as Player $P$, and, 
similarly, $x_e(t) \in \mathbb{R}^{n_e}$ is the state vector  at time $t$ of the other player, referred to as Player $E$.
The initial states $x^0_{p}$ and $x^0_{e}$ are given.
In (\ref{eq:p_dyn}), $u(t) \in U \subseteq \mathbb{R}^m$ is the control input of Player $P$, and in (\ref{eq:e_dyn}), $v(t) \in V \subseteq \mathbb{R}^\ell$ is the control input of Player $E$. 
The state vector 
$x_w(t) \in \mathbb{R}^{n_w}$ defines the game environment (or world) whose evolution is controlled by the vector of (constant) model parameters $\rho \in \mathbb{R}^k$.

The information asymmetry in the proposed differential game arises from the difference in the players' knowledge regarding the model parameter vector $\rho$. 
Without loss of generality, it is assumed that Player $E$ knows the true value of the parameter vector
$\rho$, whereas
Player $P$ knows only the nominal value of the parameter vector, denoted by $\bar{\rho}$.
It is assumed that Player $E$ also knows the nominal value $\bar{\rho}$, and is aware of the fact that Player $P$ knows only the nominal value.
Furthermore, it is assumed that at each time $t\in [t_0,t_f]$, both players have accurate knowledge of the states $x_p(t)$ and $x_e(t)$ which they can use to implement their corresponding strategies.
The initial value problem (IVP) in (\ref{eq:env_dyn}) is solved by Player $E$ to obtain the true state $x_w(t)$ while the IVP is solved by Player $P$ under nominal conditions (e.g., $\bar{\rho}$) to obtain the nominal state $\bar{x}_w(t)$.

To summarize, the set $I_E(t) = \{x_p(t),x_e(t),x_w^0,\bar{\rho},\rho\}$ is the information set of Player $E$ at time $t$ and,
similarly, $I_P(t) = \{x_p(t),x_e(t),x_w^0,\bar{\rho}\}$ is the information set of Player $P$.
Note that $I_P(t) \subset I_E(t)$, which indicates the fact that Player $E$ is more informed. 

Next, and for the sake of brevity, the combined state vector that defines the game is denoted as $x = [x_p^\top~~x_e^\top~~x_w^\top]^\top \in \mathbb{R}^n$, where $n = n_p + n_e + n_w$, and the dynamical equations in (\ref{eq:p_dyn})-(\ref{eq:env_dyn}) can be condensed as
\begin{align}
    \dot{x}(t) = f(x(t),\rho,u(t),v(t),t), \quad x(t_0) = x^0. \label{eq:gen_dyn}
\end{align}
It is assumed that the function $f$ is continuously differentiable in $(x,\rho)$, locally Lipschitz continuous in $(u,v)$ and piecewise continuous in $t$.
The \emph{original goal} of Player $P$ is to find an optimal feedback strategy 
$u[I_P(t)]$
from the admissible set 
$\mathcal{U} = \{ u : \mathbb{R}^{n_p}\times\mathbb{R}^{n_e}\times\mathbb{R}^{n_w}\times\mathbb{R}^k \rightarrow U,~\text{locally Lipschitz in } (x_p,x_e,x_w) \}$
to
minimize the payoff function 
\begin{align}
\mathcal{J}_\rho(u,v) = \phi(x(t_f),t_f) + \int_{t_0}^{t_f}L(x(t),u(t),v(t),t) \, \text{d}t, \label{eq:payoff}
\end{align}
while satisfying $q$-number of trajectory constraints of the form
\begin{align}
    &g_\rho(x_p(t),x_w(t),t) \leq 0, \label{eq:u_constr}
\end{align}
where $\phi(\cdot)$ is the terminal cost function, and $L(\cdot)$ is the running cost.
The individual components of the constraint vector function in (\ref{eq:u_constr}) are denoted using subscripts, i.e., $g_\rho = [g_{\rho1},g_{\rho2},\dots,g_{\rho q}]^\top$.

Similarly, Player $E$ wishes to find an optimal feedback strategy 
$v[I_E(t)]$
from the admissible set $\mathcal{V} = 
\{ v :\mathbb{R}^{n_p}\times\mathbb{R}^{n_e}\times\mathbb{R}^{n_w}\times\mathbb{R}^k\times\mathbb{R}^k \rightarrow V,~\text{locally Lipschitz in } (x_p,x_e,x_w) \}$
to maximize the payoff function in (\ref{eq:payoff}), while satisfying $d$-number of constraints 
\begin{align}
    &h_\rho(x_e(t),x_w(t),t) \leq 0. \label{eq:v_constr}
\end{align}
For a given choice of the functions $u$ and $v$ (these could be uninformative open-loop controllers or,  feedback strategies),
the functions $\mathcal{J}_{\rho}(\cdot)$, $g_{\rho}(\cdot)$, and $h_{\rho}(\cdot)$ can be evaluated only when the value of $\rho$ is provided, and this implicit dependency is indicated using the subscript.
The trajectory constraints in (\ref{eq:u_constr}) and (\ref{eq:v_constr}) are orthogonal state constraints (or player-specific constraints). 
Orthogonality here implies that $g_{\rho}(\cdot)$ is dependent only on the choice of control input $u(t)$ through $x_p$, and $h_{\rho}(\cdot)$ depend only on the choice of control input $v(t)$ through $x_e$.
Orthogonal state constraints are relevant to pursuit-evasion games involving obstacles, where the collision constraints are dependent only on the control choices of the individual players~\cite{Altman2009, makkapati2022desensi}.

In order to select $u$ such that the constraint in (\ref{eq:u_constr}) is satisfied, Player $P$ requires exact information regarding the environment.
However, he only knows the nominal value $\bar{\rho}$ and, as a result, he cannot ensure meeting his trajectory constraint.
It is also assumed that Player $P$ realizes the violation of constraint in (\ref{eq:u_constr}) only when it occurs, resulting in the game termination. 
Consequently, the asymmetric information game ends when the following condition is met
\begin{align}
    &\Psi(x(t_f),t_f) = 0 \text{ or } g_{{\rho}}(x_p(t_f),x_w(t_f),t_f) \nleq 0 \text{ or } h_{{\rho}}(x_e(t_f),x_w(t_f),t_f) \nleq 0, \label{eq:ter_cond}
\end{align}
where $\Psi(x(t_f),t_f) = 0$ is a function that captures the terminal condition of the game.
It is assumed that the mappings $f:\mathbb{R}^n\times\mathbb{R}^k\times U \times V \times \mathbb{R}\rightarrow \mathbb{R}^n$, $\mathcal{J}_\rho: \mathcal{U} \times \mathcal{V} \rightarrow \mathbb{R}$, $g_\rho: \mathbb{R}^{n_p} \times \mathbb{R}^{n_w} \times \mathbb{R} \rightarrow \mathbb{R}^q$, $h_\rho: \mathbb{R}^{n_e} \times \mathbb{R}^{n_w} \times \mathbb{R} \rightarrow \mathbb{R}^{d}$, $\Psi: \mathbb{R}^{n} \times \mathbb{R} \rightarrow \mathbb{R}$, are common knowledge for both players.

For the aforementioned asymmetric differential game formulation, we first develop deceptive strategies for Player $E$ in Section \ref{sec:decept}.
In Section \ref{sec:cdp},
we propose risk-sensitive strategies for Player $P$ that provide safer trajectories, thereby reducing the chance of constraint violation while optimizing his original payoff function. 
It is assumed that the proposed deviations from the original payoff function in 
Sections~\ref{sec:decept} and \ref{sec:cdp} by the individual players are not known to their respective adversaries.
In other words, the mapping $\mathcal{D}_\rho: \mathcal{U} \times \mathcal{V} \rightarrow \mathbb{R}$ (defined in Section \ref{sec:decept}) is known only to Player $E$, and the mapping $\mathcal{R}_\rho: \mathcal{U} \rightarrow \mathbb{R}$ (defined in Section \ref{sec:cdp}) is known only to Player $P$. 

\section{The Deceptive Agent (Player E)}
\label{sec:decept}

From the discussion in the previous section, it is evident that Player $E$ is the more informed player among the two agents.
Given the information structure in Section~\ref{sec:prelims}, 
the optimal feedback strategy $v^*_o$ that serves the original goal of Player $E$ is obtained by solving the differential game
\begin{equation}\tag{$\mathcal{E}_o$}
	v^*_o = \underset{\substack{v \in \mathcal{V},~h_{{\rho}}\leq 0}}{\arg\max}~ \mathcal{J}_{{\rho}}\left(\hat{u}(v),~v\right),
\end{equation}
where,
\begin{equation}
	\hat{u}(v) = \underset{\substack{u \in \mathcal{U},~g_{\bar{\rho}}\leq 0}}{\arg\min}~\mathcal{J}_{\bar{\rho}}(u,v) \label{eq:uhat}
\end{equation}
is the optimal strategy of Player $P$ as a function of Player $E$'s control $v$.
In \eqref{eq:uhat} the subscript $\bar{\rho}$ in the $\arg\min$ indicates that Player $E$ is aware of the fact that Player $P$ knows only the nominal value of the model parameters.
Therefore, Player $E$ can estimate Player $P$'s optimal strategy, as a function of $v$, that serves Player $P$'s original goal against the nominal model.

In this paper, deception is modeled as a game where the more informed player achieves its alternative/deceptive goal by \emph{misleading} the less informed player---in our case Player $P$.
Since Player $P$ follows a partial-state feedback strategy by observing only the state $x_e$ and not $x_w$, Player $E$ can influence the trajectory of Player $P$.
For example, Section \ref{sec:pe} defines a two-player PE game with a dynamic obstacle, where the original goal of the evader (more informed player) is to maximize capture time.
The evader can also employ a deceptive goal of leading the pursuer (less informed player) towards the obstacle since it is assumed that the pursuer cannot observe the true position of the obstacle.

The pursuer's lack of knowledge regarding the speed and location of the obstacle motivates the evader to consider an alternative goal, namely, to misguide the pursuer so that he collides with the obstacle.
To summarize, in order to deceive, the more informed player considers two components: a) her adversary's feedback strategy; and b) a deceptive goal.
Accordingly, Player $E$ computes an optimal deceptive strategy $v^*_d$ by solving the game  
\begin{equation}\tag{$\mathcal{E}_d$}
v^*_d = \underset{\substack{v \in \mathcal{V},~h_{{\rho}}\leq 0}}{\arg\max}~ \alpha_o \mathcal{J}_{{\rho}}\big(\hat{u}(v),~v\big) + \alpha_d \mathcal{D}_{{\rho}}\big(\hat{u}(v),~v\big),
\end{equation}
where $\mathcal{D}_{{\rho}}\big(\hat{u}(v),~v\big)$ denotes the deception payoff that Player $E$ wants to maximize along with the original payoff function.
The positive weights $\alpha_o, \alpha_d \in \mathbb{R}$ balance the individual components of the multi-objective payoff function in ($\mathcal{E}_d$).
The instance where the evader does not consider its original payoff function (i.e., $\alpha_o = 0$) can be termed as \emph{pure deception}, which is analyzed for the proposed PE game in Section \ref{sec:pe}.
In this work, the deceptive goal of Player $E$ is assumed to be the violation of Player $P$'s trajectory constraints in (\ref{eq:u_constr}), i.e.,   
\begin{align}
    \mathcal{D}_{{\rho}}\big(\hat{u}(v),~v\big) = \sum_{i=1}^q g_{{\rho i}}(x_p(t),x_w(t),t),
\end{align}
where $g_{{\rho i}}(x_p(t),x_w(t),t)$ denotes the $i^{th}$ component of the constraint function in (\ref{eq:u_constr}).

It has to be noted that in order to develop deceptive strategies, it is not necessary for Player $E$ to consider the optimal strategy of Player $P$ given in (\ref{eq:uhat}).
The more informed player (Player $E$ in this case) may consider any  
sub-optimal feedback strategy for Player $P$.
In the following section, we discuss strategies for the less informed player, which are developed using the theory of sensitivity functions that are appropriate for the proposed differential game formulation.

\section{The Risk-Averse Agent (Player P)}
\label{sec:cdp}

In this section, using sensitivity functions, we first capture the variations in the pursuer's constraint function (\ref{eq:u_constr}) under parametric uncertainty.
The constraint variations are then appropriately weighted using a relevance function to obtain the \emph{relevant constraint sensitivity} (RCS), which is used to construct a regularizer that captures the risk of constraint violation.
The following section introduces the theory of sensitivity functions, which provides first-order estimates of the effect of parameter variations on the trajectories of the proposed differential game.

\subsection{Sensitivity Functions}

Consider the dynamics in (\ref{eq:gen_dyn}), and assume variations in the model parameter vector $\rho \in \mathbb{R}^k$, with $\bar{\rho}$ being its nominal value.
As specified in Section \ref{sec:prelims}, assume that $f(x,\rho,u,v,t)$ is continuously differentiable with respect to $x$ and $\rho$ for all $(u,v,t) \in U \times V \times [t_0,t_f]$.
The solution to the differential equation (\ref{eq:gen_dyn}) starting from the initial condition $x(t_0) = x^0$ using admissible open-loop controllers is given by
\begin{align}
x(\rho,t) = x^0 + \int_{t_0}^{t} f(x(\rho,\tau),\rho,u(\tau),v(\tau),\tau) \, \text{d}\tau.
\end{align}
Since $f(x,\rho,u,v,t)$ is differentiable with respect to $\rho$, using the Leibniz integral rule for differentiation under the integral, it follows that
\begin{align}
\frac{\partial x}{\partial \rho}(\rho,t) &= \int_{t_0}^{t} \left[ \frac{\partial f}{\partial x}(x(\rho,\tau),\rho,u(\tau),v(\tau),\tau) \frac{\partial x}{\partial \rho}(\rho,\tau)\right. \nonumber\\
&~~~~~~~~~~~~~\left.+ \frac{\partial f}{\partial \rho}(x(\rho,\tau),\rho,u(\tau),v(\tau),\tau) \right]  \mathrm{d}\tau. \label{eq:LIR}
\end{align}
The \emph{parameter sensitivity function} $S:[t_0,t_f]\rightarrow \mathbb{R}^{n\times\ell}$ is obtained by evaluating (\ref{eq:LIR}) at the nominal conditions ($\rho = \bar{\rho}$) as
\begin{align}
S(t) = \dfrac{\partial x (\rho,t)}{\partial \rho}\bigg\rvert_{x = x(\bar{\rho},t)}. \label{eq:SF}
\end{align}
From (\ref{eq:LIR}), the dynamics of the sensitivity function, defined in (\ref{eq:SF}), can be obtained using the fundamental theorem of calculus \cite{courant1965introduction} as
\begin{align}
\dot{S}(t) = A(t)S(t) + B(t), \quad S(t_0) = 0_{n \times \ell}, \label{eq:sensi_eq}
\end{align}
where,
\begin{align}
A(t) &= \dfrac{\partial f(x,\rho,u(t),v(t),t)}{\partial x}\bigg\rvert_{x = x(\bar{\rho},t),\, \rho = \bar{\rho}}, \quad B(t) &= \dfrac{\partial f(x,\rho,u(t),v(t),t)}{\partial \rho}\bigg\rvert_{x = x(\bar{\rho},t),\, \rho = \bar{\rho}}.
\end{align}
The initial condition for the sensitivity function is the zero matrix Since the initial state is given (fixed).
Equation (\ref{eq:sensi_eq}) is called the \emph{sensitivity equation} in the literature \cite{khalil}.
To compute the sensitivity function over time, the state $x$ has to be propagated using the dynamics in (\ref{eq:gen_dyn}) under the nominal conditions,
\begin{align}
\dot{\bar{x}} = f(\bar{x},\bar{\rho},u,v,t), \quad \bar{x}(t_0) = x^0. \label{eq:x_dynamics}
\end{align}
Note that Player $P$ has information only about the nominal value of the parameter.
Here, $\bar{x}(t) = x(\bar{\rho},t)$ denotes the nominal state at time $t$, as computed by Player $P$ while solving the differential equation in (\ref{eq:x_dynamics}), given the controllers $u$ and $v$.
From the properties of continuous dependence with respect to the parameters and the differentiability of solutions of ordinary differential equations \cite{courant1965introduction}, 
for sufficiently small variations from the nominal value $\bar{\rho}$, the solution $x(\rho,t)$ can be approximated by
\begin{align}
x(\rho,t) \approx x(\bar{\rho},t) + S(t)(\rho - \bar{\rho}). \label{eq:sensi_approx}
\end{align}
This is a first-order approximation of $x(\rho,t)$ about the nominal solution $x(\bar{\rho},t)$.
In the next section, we develop a scheme to generate safe trajectories for Player $P$ by penalizing a risk estimate that is defined using sensitivity functions.

\subsection{Relevant Constraint Sensitivity}
For the asymmetric differential game in Section \ref{sec:prelims}, assuming the constraint function $g_\rho(\cdot)$ is a smooth function in $x$, we obtain the sensitivity of the constraint function as
\begin{align}
    S_g(t) = \diffp{g_\rho}{\rho}\bigg\rvert_{\rho = \bar{\rho}} 
    &= \left(\diffp{g_\rho}{x}\dfrac{\partial x (\rho,t)}{\partial \rho} \right)\bigg\rvert_{x = \hat{x}(t),\rho = \bar{\rho}} \nonumber \\
    &= \left(\diffp{g_\rho}{x}S(t) \right)\bigg\rvert_{x = \hat{x}(t),\rho = \bar{\rho}}.
 \end{align}
In \cite{sarabu2020safe}, it has been argued that ``variations in the constraint value when the system is far from the constraint boundary are not as important as when the system is close to the constraint boundary."
Therefore, to account for the fact that the constraint variations are more likely to cause constraint violations when the system is closer to the constraint boundary, we introduce a \emph{relevance function} $\gamma:\mathbb{R} \rightarrow [0,\infty)$, which has the form
\begin{align}
    \gamma(z) = \begin{cases}
    \tilde{\gamma}(z), \quad \text{if } z\leq 0,\\
    \tilde{\gamma}(0), \quad \text{if } z > 0,
    \end{cases}
\end{align}
where $\tilde{\gamma}:\mathbb{R} \rightarrow [0,\infty)$ is a continuous function that is monotonically increasing over the interval $(-\infty,0]$, i.e.,   $\tilde{\gamma}(z) \geq \tilde{\gamma}(y)$, if $z > y$ for all $z,y \le 0$ \cite{sarabu2020safe}. 
Examples of $\tilde{\gamma}(z)$ include $e^{-z^2}$ (Gaussian), $\text{max}(0,1-|z|)$ (Hat function), $1/(1+z^2)$, etc.

Next, we construct the \emph{relevant constraint sensitivity} (RCS) matrix $S_\gamma:[0,\infty) \rightarrow \mathbb{R}^{q\times k}$ as
\begin{align}
    S_\gamma(t) = RS_g(t),
\end{align}
where $R = \text{diag}\left(\gamma(g_{\bar{\rho}1}),\gamma(g_{\bar{\rho}2}),\dots,\gamma(g_{\bar{\rho}q})\right)$.
For the purpose of the analysis, as in  \cite{sarabu2020safe},  
the derivative of the logistic function $s(z) = 1/(1 + e^{-z})$ is chosen as the candidate relevance function, that is,
\begin{align}
    \tilde{\gamma}(z) = s(z)(1 - s(z)).
\end{align}
The sensitivity matrix $S_\gamma$ captures the intuition that more importance to variations should be given near the constraint boundary.
The proposed risk estimate for Player $P$ can be obtained as
\begin{align}
    \mathcal{R}_{\bar{\rho}}(u) = \int_{t_0}^{t_f} \|\vecop S_\gamma(t)\|_{Q}^{2} \, \text{d}t, \label{eq:risk_measure}
\end{align}
where $\|w\|_Q = w^\top Q w$, and $Q \succeq 0$.
Here, $\vecop(\cdot)$ denotes the standard vectorization operator that converts a $\mathbb{R}^{q\times k}$ matrix into $qk$-dimensional vector.
The risk estimate in (\ref{eq:risk_measure}) is also referred to as the RCS regularizer, and can be understood as a sensitivity-based measure that captures the risk of constraint violation \cite{sarabu2020safe}.
Using the proposed risk estimate in (\ref{eq:risk_measure}), a risk-averse strategy for Player $P$ is proposed in the following section.

\subsection{A Risk-Averse Strategy for the Pursuer}

Since Player $P$ does not know the true value of the parameter vector, he can solve the game only under nominal conditions.
Therefore, the optimal feedback strategy $u^*_o$ that serves the original goal of Player $P$ is obtained by solving the differential game
\begin{equation}\tag{$\mathcal{P}_o$}
u^*_o = \underset{\substack{u \in \mathcal{U},~g_{\bar{\rho}}\leq 0}}{\arg\min}~ \mathcal{J}_{\bar{\rho}}\left(u,~\hat{v}(u)\right),
\end{equation}
where,
\begin{align}
  \hat{v}(u) =  \underset{\substack{v \in \mathcal{V},~h_{\bar{\rho}}\leq 0}}{\arg\max}~\mathcal{J}_{\bar{\rho}}(u,v). \label{eq:e_os_nom}
\end{align}

In order to construct safe trajectories (that minimize the chance of constraint violation) for the minimizing player under parametric uncertainties, we propose a modified payoff function of the form
\begin{align}
    \tilde{\mathcal{J}}_{\bar{\rho}}(u,v) = \mathcal{J}_{\bar{\rho}}(u,v) + \mathcal{R}_{\bar{\rho}}(u).  \label{eq:CDP_cost}
\end{align}
With $\hat{v}(u)$ defined in (\ref{eq:e_os_nom}), the optimal desensitized, or risk-averse, strategy $u^*_r$ for Player $P$ can be defined as
\begin{equation}\tag{$\mathcal{P}_r$}
u^*_r = \underset{\substack{u \in \mathcal{U},~g_{\bar{\rho}}\leq 0}}{\arg\min}~ \tilde{\mathcal{J}}_{\bar{\rho}}\left(u,~\hat{v}(u)\right).
\end{equation}
It can be observed that the games $\mathcal{P}_o$ and $\mathcal{P}_r$ are equivalent when $Q = 0$ as per (\ref{eq:risk_measure}).

In the following section, the proposed deceptive strategy ($v^*_d$) for Player $E$ and risk-averse strategy ($u^*_r$) for Player $P$ are analyzed for a two-player pursuit-evasion problem with an uncertain dynamic obstacle. 


\section{Pursuit-Evasion with Uncertain Dynamic Obstacle} 
\label{sec:pe}
In this section, Player $P$ is considered to be the pursuer, $x_p(t) \in \mathbb{R}^2$ denotes the pursuer's position, and $u(t) \in \{y\in\mathbb{R}^2:\|y\|_2 = u_c\}$ denotes his velocity vector.
Similarly, Player $E$ is considered to be the evader, $x_e(t) \in \mathbb{R}^2$ denotes the evader's position, and $v(t) \in \{y\in\mathbb{R}^2:\|y\|_2 = v_c\}$ denotes her velocity vector.
The individual components for the two-dimensional vectors in this section are denoted using the subscripts 1 and 2,
that is, $x_p = [x_{p1}, x_{p2}]^\top$.
Note that the pursuer and the evader are assumed to be moving at constant speeds $u_c$ and $v_c$, respectively, with heading control.
To ensure the capturability of the evader, it is assumed that $u_c > v_c$.
The pursuer's objective is to capture the evader by entering its capture zone, assumed here to be a disk of radius $\epsilon > 0$ centered at the instantaneous position of the evader,
whereas the evader's objective is to avoid capture indefinitely.
Therefore, the original payoff function the pursuer wants to minimize, while the evader wants to maximize, can be written as
\begin{align}
    \mathcal{J}_\rho(u,v) = t_{\text{cap}}, \label{eq:tc_cost}
\end{align}
where $t_{\text{cap}}$ is the time to capture.

The proposed PE game environment contains a single dynamic circular obstacle of radius $r_o$, where $x_w \in \mathbb{R}^2$ denotes the position of the obstacle's center.
The game dynamics, as per (\ref{eq:p_dyn})-(\ref{eq:env_dyn}), can be expressed as
\begin{align}
    \dot{x}_p(t) = u(t), \quad \dot{x}_e(t) = v(t), \quad \dot{x}_w(t) = \rho. \label{eq:pegame_dyn}
\end{align}
From (\ref{eq:pegame_dyn}), it is evident that the uncertain parameter in the proposed game is the velocity of the obstacle.
Both players have to avoid colliding with the obstacle while they traverse the environment to achieve their 
respective objectives.
Subsequently, the trajectory constraint functions $g_\rho(\cdot)$ and $h_\rho(\cdot)$ take the form
\begin{align}
    g_\rho &= r_o^2 - \|x_p(t) - x_w(t)\|_2^2 \leq 0, \label{eq:p_constraint}\\
    h_\rho &= r_o^2 - \|x_e(t) - x_w(t)\|_2^2 \leq 0. \label{eq:e_constraint}
\end{align}

The evader knows the true velocity of the obstacle (${\rho}$), and the pursuer only knows the nominal value ($\bar{\rho}$).
In this game, the pursuer cannot observe the obstacle's movements until the pursuer collides with the obstacle.
As mentioned earlier, the original objective of the evader is to maximize the time to capture. 
Since the game ends if the pursuer hits the obstacle or captures the evader, and since the pursuer is unaware of the obstacle's exact whereabouts, it can be argued that the evader is motivated to drive the pursuer so as to collide with the obstacle in order to avoid capture.

In order to address the uncertainty in the obstacle's velocity, safe trajectories for the pursuer are generated using the regularizer introduced in Section \ref{sec:cdp}.
Since the obstacle has simple dynamics, a closed-form expression of the sensitivity of the pursuer's constraint function $g_\rho$ with respect to the uncertain parameter $\rho$ at time $t$ is given by
\begin{align}
    S_g(t) = \left[\begin{array}{cc}
         \dfrac{\partial g_\rho}{\partial \rho_1}\\
         \\
         \dfrac{\partial g_\rho}{\partial \rho_2} 
    \end{array}\right]^\top = 2t\left[\begin{array}{cc}
        x_{p1}(t) - x_{w1}(t)\\
         \\
        x_{p2}(t) - x_{w2}(t)
    \end{array}\right]^\top. \label{eq:pe_Sg}
\end{align}

In order to construct feedback strategies, it is assumed that both players know the initial position of the obstacle, and can observe the instantaneous positions of themselves and each other for all time. 
For the rest of the game, the pursuer evaluates only the nominal position of the obstacle using the nominal value of the velocity vector ($\bar{\rho}$) and the evader observes the true position of the obstacle for all time.
Given the original payoff function (\ref{eq:tc_cost}), the constraint function (\ref{eq:p_constraint}), and the constraint sensitivity function (\ref{eq:pe_Sg}), obtaining the optimal feedback strategies of the players for the games $\mathcal{E}_o$, $\mathcal{P}_o$, $\mathcal{E}_d$, and $\mathcal{P}_r$, defined in Sections \ref{sec:decept} and \ref{sec:cdp}, in closed form is elusive. 
The games are therefore solved numerically in a receding horizon fashion as illustrated in the following section.

\subsection{Receding Horizon Control in Discrete Time}

In order to construct a receding horizon control (RHC) law for both players, we rewrite the dynamics (\ref{eq:pegame_dyn}) in discrete time as follows.
\begin{align}
    &x^{k+1}_p = x^{k}_p + u^{k} \Delta t,\\
    &x^{k+1}_e = x^{k}_e + v^{k} \Delta t,\\
    &x^{k+1}_w = x^{k}_w + \rho \Delta t,
\end{align}
where the superscript $k$ denotes the variable at the $k^{th}$ time-step, and $\Delta t$ is the length of time interval considered for discretization (assumed to be a constant).
At each time step, let $N\Delta t$ be the length of the horizon over which the games are solved to obtain the open-loop controls of both players.

Since the players follow an RHC law with a fixed time horizon, the original payoff function in (\ref{eq:tc_cost}), which is the time-to-capture, is updated to be the relative distance between the players at the end of the horizon.
Consequently, at the $k^{th}$ time-step in the RHC framework, the original payoff function that the players want to optimize is given by
\begin{align}
    &\mathcal{J}_\rho(\check{u}^{1:N},\check{v}^{1:N}) = \|\check{x}_p^{k+N} - \check{x}_e^{k+N}\|_2,
\end{align}
where $\check{u}^{1:N},~\check{v}^{1:N}$ are the predicted control inputs of the players along the receding horizon.
Similarly, $\check{x}_p^{k+i}$, $\check{x}_e^{k+i}$, $i \in \{1,\dots,N\}$ are the players' predicted position vectors along the receding horizon, obtained using the control inputs $\check{u}^{1:N},~\check{v}^{1:N}$.
Note that the time interval $\Delta t$ is assumed to be constant and hence, the relevant constraint sensitivity-based risk estimate for the pursuer at the $k^{th}$ time-step is given by 
\begin{align}
    \mathcal{R}_{\bar{\rho}}(\check{u}^{1:N}) = \sum_{i = 1}^{N}\|\vecop \check{S}_\gamma^{k+i}\|_{Q}^{2},
\end{align}
where $\check{S}_\gamma^{k+i}$ is the RCS at the $i^{th}$ step in the receding horizon. 

For the games $\mathcal{P}_o$, $\mathcal{P}_r$, $\mathcal{E}_o$, the pursuer and the evader compute their respective RHC-based inputs at the $k^{th}$ time-step using Algorithm \ref{algo:rhc_minmax}.
The iterative computation to obtain the fixed points in the aforementioned games in Algorithm \ref{algo:rhc_minmax} is popularly known as the Gauss-Seidel method \cite{li1987distributed}.
The constraints $g_\rho$ and $h_\rho$ are imposed at discrete time-steps in the optimization problems stated in the algorithm. 

\begin{algorithm}
	\caption{Receding horizon control law for discrete game}\label{algo:rhc_minmax}
	\begin{algorithmic}[1]
		\Procedure{Pursuer\_Game}{$N$, $\Delta t$, $x^k_p$, $x^k_e$} \Comment{($\mathcal{P}_o$/$\mathcal{P}_r)$}
		\State Initialize with $[\check{v}^{1:N}]_0$, $j\leftarrow 1$ 
		\Repeat 
		\State $[\check{u}^{1:N}]_j \leftarrow \underset{\check{u}^{1:N},~g_{\bar{\rho}}\leq 0}{\arg\min}~\tilde{\mathcal{J}}_{\bar{\rho}}(\check{u}^{1:N},[\check{v}^{1:N}]_{j-1})$
		\State $[\check{v}^{1:N}]_j \leftarrow \underset{\check{v}^{1:N},~h_{\bar{\rho}}\leq 0}{\arg\max}~\mathcal{J}_{\bar{\rho}}([\check{u}^{1:N}]_j,\check{v}^{1:N})$
		\State $j \leftarrow j+1$
		\Until{convergence}
		\State $u^k \leftarrow [\check{u}^1]_{j-1}$
		\State \Return $u^k$
		\EndProcedure
            \Procedure{Evader\_Game}{$N$, $\Delta t$, $x^k_p$, $x^k_e$} \Comment{($\mathcal{E}_o)$}
		\State Initialize with $[\check{v}^{1:N}]_0$, $j\leftarrow 1$ 
		\Repeat 
		\State $[\check{u}^{1:N}]_j \leftarrow \underset{\check{u}^{1:N},~g_{\bar{\rho}}\leq 0}{\arg\min}~\mathcal{J}_{\bar{\rho}}(\check{u}^{1:N},[\check{v}^{1:N}]_{j-1})$
		\State $[\check{v}^{1:N}]_j \leftarrow \underset{\check{v}^{1:N},~h_{{\rho}}\leq 0}{\arg\max}~\mathcal{J}_{{\rho}}([\check{u}^{1:N}]_j,\check{v}^{1:N})$
		\State $j \leftarrow j+1$
		\Until{convergence}
		\State $v^k \leftarrow [\check{v}^1]_{j-1}$
		\State \Return $v^k$
		\EndProcedure
	\end{algorithmic}
\end{algorithm}

The proposed discrete-time deception payoff for the evader in the RHC framework is given by 
\begin{align}
    \mathcal{D}_{{\rho}}(\hat{u}(\check{v}^{1:N}),\check{v}^{1:N}) = -\|\hat{x}_p^{k+N} - x_w^{k+N}\|_2,
\end{align}
where $\hat{x}^{k+i}_p$, $i \in \{1,\dots,N\}$ is the pursuer's position along the receding horizon that is obtained using the feedback strategy $\hat{u}$, as discussed in Section \ref{sec:decept}.
In order to compute the deceptive strategy for the evader, as per the game defined in $\mathcal{E}_d$, we assume the pursuer's feedback strategy to be pure pursuit. 
Consequently,
\begin{align}
    \hat{u}^i = u_c\frac{\check{x}^{k+i-1}_e - \hat{x}^{k+i-1}_p}{\|\check{x}^{k+i-1}_e - \hat{x}^{k+i-1}_p\|_2}, \quad i \in \{1,\dots,N\}.
\end{align}

In the following section, we present simulation results for the proposed pursuit-evasion game with the players employing desensitized and deceptive strategies.
The simulations were performed in MATLAB using its built-in function \texttt{fmincon} in conjunction with YALMIP \cite{Lofberg2004} for the optimization problems shown in Algorithm \ref{algo:rhc_minmax}. 
Note that $[\,.\,]_j$ indicates the $N$-step control input obtained at the $j^{th}$ iteration of the chosen Gauss-Seidel method of optimization in Algorithm \ref{algo:rhc_minmax}.
The convergence criterion to obtain the equilibrium points for the games $\mathcal{P}_o/\mathcal{P}_r$ and $\mathcal{E}_o$, as per the Algorithm \ref{algo:rhc_minmax}, is chosen to be $\|[\check{u}^{1:N}]_{j}-[\check{u}^{1:N}]_{j-1}\|,\|[\check{v}^{1:N}]_{j}-[\check{v}^{1:N}]_{j-1}\|\leq 5\times10^{-3}$.

Note that the analyses presented in this paper are based on heuristic arguments, and in this regard the paper assumes the existence of fixed points for the games $\mathcal{P}_o$, $\mathcal{P}_r$, $\mathcal{E}_o$.
It should be stressed at this point that although we do not yet have a formal proof of the existence of a fixed point, in all our numerical simulations Algorithm~1 converged for all simulations conducted in this paper.

\subsection{RCS Fields}
\label{subsec:rcs_fields}

\begin{figure}[htb!]
    \centering
    \subfigure[Both the components $\rho_1$ and $\rho_2$ are uncertain]{\includegraphics[width = 0.4\textwidth]{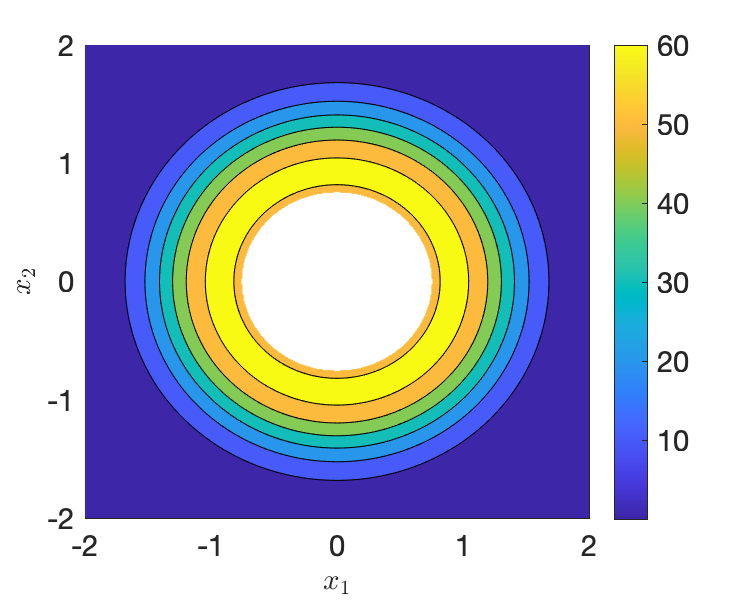}}
    \subfigure[Only the component $\rho_1$ is uncertain]{\includegraphics[width = 0.4\textwidth]{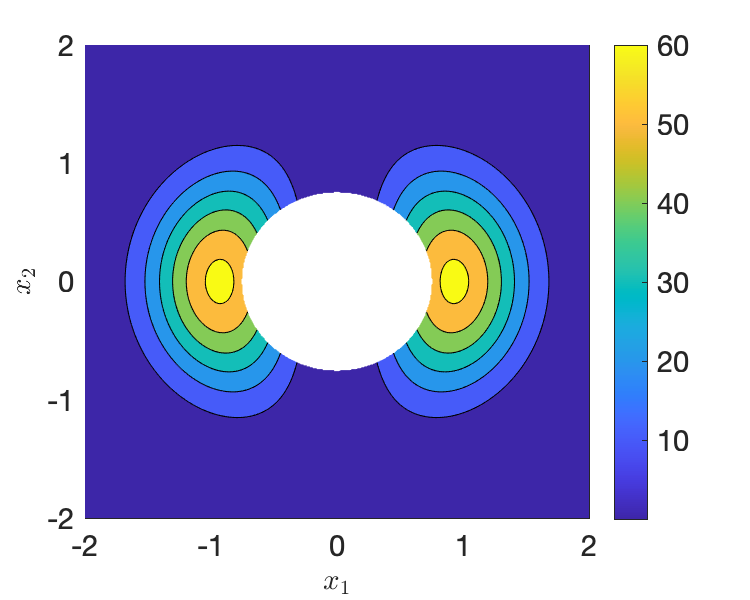}}
    \caption{Norm of the proposed RCS ($S_\gamma$) matrix around the evader at $t=10$ for two different instances of uncertainty in obstacle' velocity}
    \label{fig:RCS_norm}
    \vspace*{-5pt}
\end{figure}

In this section, the pursuer's desensitized strategy is pitted against the evader's deceptive and non-deceptive strategies.
To this end, the effects of desensitization are studied by first analyzing the RCS regularizer. 
Figure \ref{fig:RCS_norm} presents the 2-norm of the RCS matrix in (\ref{eq:pe_Sg}) for two different instances of uncertainty in the obstacle's velocity that are considered in this paper for simulation purposes.
For both contour plots in Fig. \ref{fig:RCS_norm}, the white circular region in the middle represents the nominal obstacle, whose center is at the origin with the radius $r_o = 0.75$. 
Figure \ref{fig:RCS_norm}(a) shows the field of RCS regularizer around the obstacle in the $x_1$-$x_2$ plane, when the two components of the obstacle's velocity $\rho_1$ and $\rho_2$ are assumed to be uncertain.
It can be observed that the RCS norm is larger closer to the obstacle, indicating that the agent would be penalized when it is closer to the obstacle and vice versa.
It can be noted that the contours of the RCS norm are concentric circles around the obstacle.
In a case where only the component $\rho_1$ is assumed to be uncertain, and since $\rho_2$ has no uncertainty, there should not be any penalty on any of the points along the $x_2$-axis, as can be seen in Fig. \ref{fig:RCS_norm}(b).
In the second case, the RCS regularizer field has two poles on the $x_1$-axis, and is symmetric about the $x_2$-axis.
As a result, the second case may lead to the agent getting into a local minimum along the axis of no uncertainty for certain simulation parameters, as can be observed in some of the upcoming simulations.
Note that the case where $\rho_1$ is certain, and $\rho_2$ is uncertain is a transformation of the second case with the poles on the $x_2$ axis.  

An alternative way of representing the obstacle's velocity vector is by using the polar coordinates $\|\rho\|_2$ (total speed) and $\psi_\rho$ (heading with respect to $x_1$-axis).
In this regard, the sensitivity of the constraint function ($g_\rho$) with respect to the components in the polar coordinate system in continuous time (analogous to the constraint sensitivity components in (\ref{eq:pe_Sg})) can be obtained as
\begin{align}
    &\dfrac{\partial g_\rho}{\partial \|\rho\|_2} = 2t\left[(x_{p2} - x_{w2})\sin \psi_\rho + (x_{p1} - x_{w1})\cos \psi_\rho\right], \label{eq:Sg_speed}\\
    &\dfrac{\partial g_\rho}{\partial \psi_\rho} = 2\|\rho\|_2t\left[(x_{p2} - x_{w2})\cos \psi_\rho - (x_{p1} - x_{w1})\sin \psi_\rho 
    \right]. \label{eq:Sg_heading}
\end{align}
The derivation of the above expressions is straightforward and is not included here for the sake of brevity. 
Also, the time dependencies of the variables in (\ref{eq:Sg_speed}) and (\ref{eq:Sg_heading}) are dropped to keep the expressions succinct.
From (\ref{eq:Sg_speed}), it can be inferred that in the instance where $\|\rho\|_2$ (total speed) alone is the uncertain parameter, the related RCS field looks similar to the one in Fig. \ref{fig:RCS_norm}(b), and the poles will lie along the nominal velocity vector.
At the same time, in the instance where $\psi_\rho$ (heading) alone is the uncertain parameter, the related RCS field the related RCS field again looks similar to the one in Fig. \ref{fig:RCS_norm}(b), but the poles will lie orthogonal to the nominal velocity vector.
The discussion in  \cite{sarabu2020safe} provides additional insights into the characteristics of the RCS field by comparing it with the contour for constraint sensitivity ($S_g$) in motion planning problems.

\begin{figure}[htb!]
    \centering
    \includegraphics[width = 0.32\textwidth]{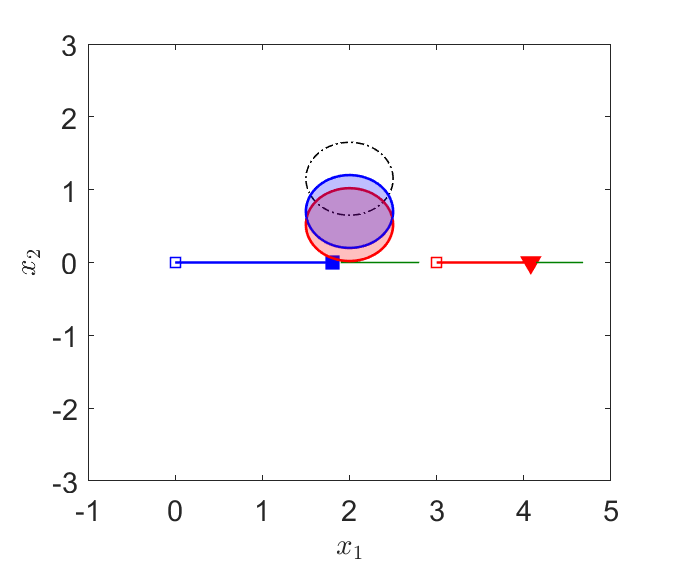}
    \caption{An instance where the pursuer does not penalize the proposed risk estimate that leads to collision (pursuer solves $\mathcal{P}_o$). 
    Black dotted circle - initial position of the obstacle; Blue solid circle - Nominal Obstacle; Red solid circle - True Obstacle; Green line - RHC-based look-ahead of the player.}
    \label{fig:DOC_colli}
    \vspace*{-5pt}
\end{figure}

\begin{figure*}[htb!]
    \centering
    \subfigure[$t = 0.4$]{\includegraphics[width = 0.32\textwidth]{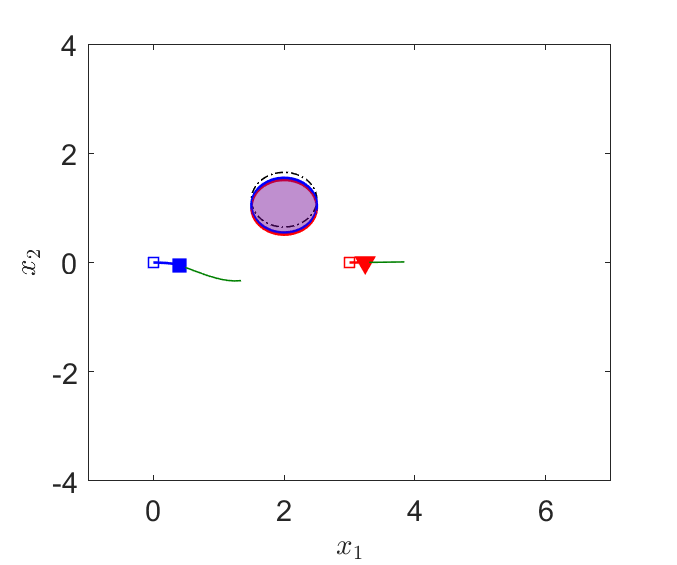}}
    \subfigure[$t = 2.8$]{\includegraphics[width = 0.32\textwidth]{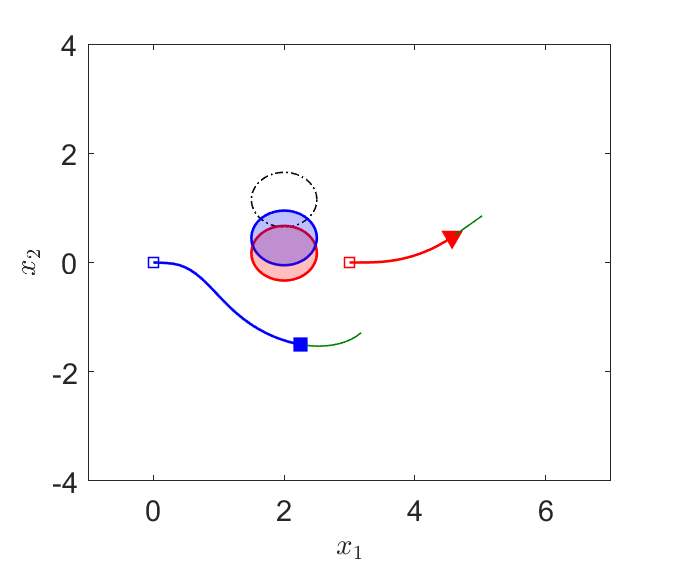}}
    \subfigure[$t = 5.7$]{\includegraphics[width = 0.32\textwidth]{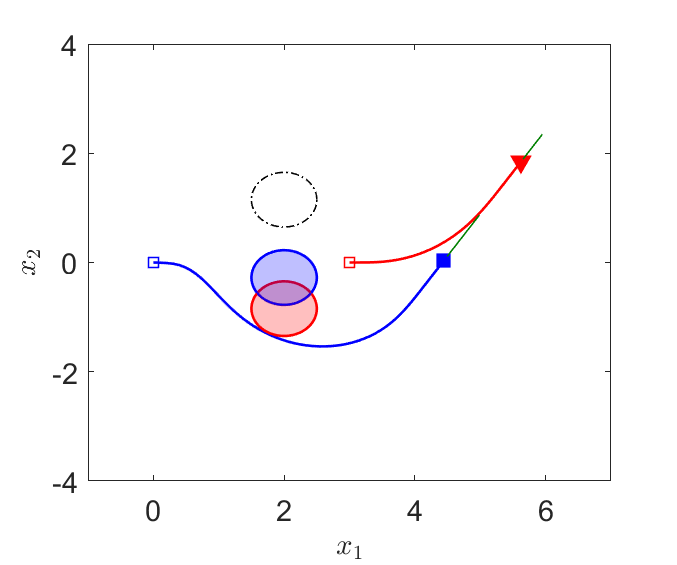}}
    \caption{Simulation results for an instance where the pursuer penalizes the proposed risk estimate with $Q=1$, and successfully captures the evader while avoiding the obstacle. The uncertain parameter is the obstacle's speed along $x_2$-axis.}
    \label{fig:Desensi_NoDecept_1}
    \vspace*{-5pt}
\end{figure*}

\begin{figure*}[htb!]
    \centering
    \subfigure[$t = 1.0$]{\includegraphics[width = 0.32\textwidth]{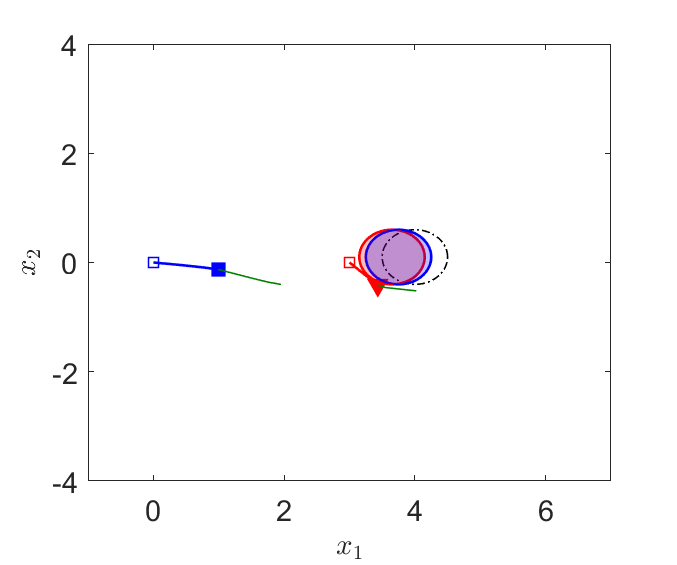}}
    \subfigure[$t = 3.0$]{\includegraphics[width = 0.32\textwidth]{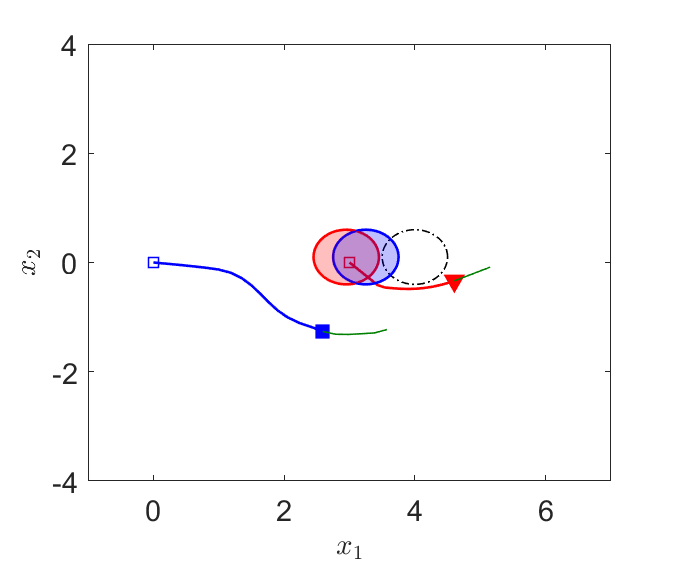}}
    \subfigure[$t = 5.6$]{\includegraphics[width = 0.32\textwidth]{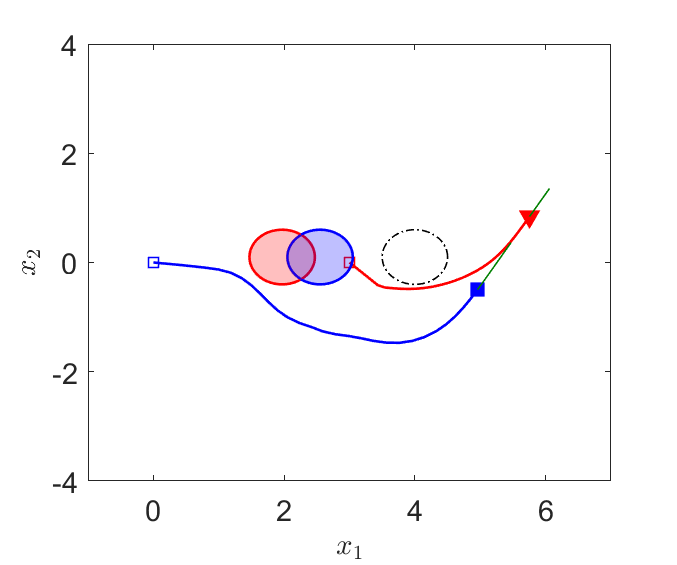}}
    \caption{Simulation results for an instance where the pursuer employs the proposed desensitized strategy with $Q=1$ while the evader \emph{interacts} with the obstacle. The uncertain parameter is the obstacle's speed along $x_1$-axis.}
    \label{fig:Desensi_NoDecept_2}
    \vspace*{-5pt}
\end{figure*}

\begin{figure*}[htb!]
    \centering
    \subfigure[$t = 2.4$]{\includegraphics[width = 0.32\textwidth]{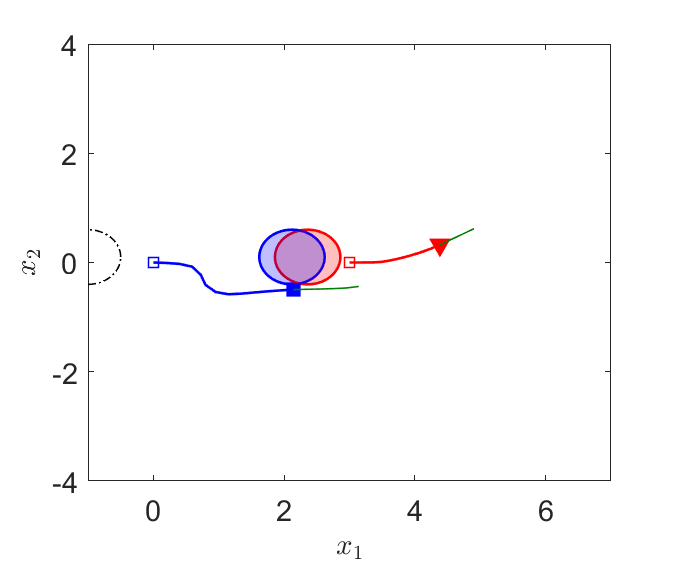}}
    \subfigure[$t = 3.2$]{\includegraphics[width = 0.32\textwidth]{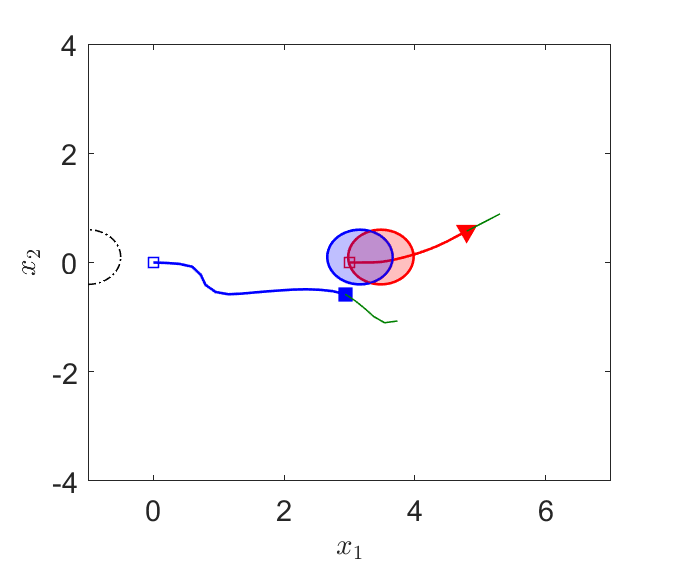}}
    \subfigure[$t = 6.4$]{\includegraphics[width = 0.32\textwidth]{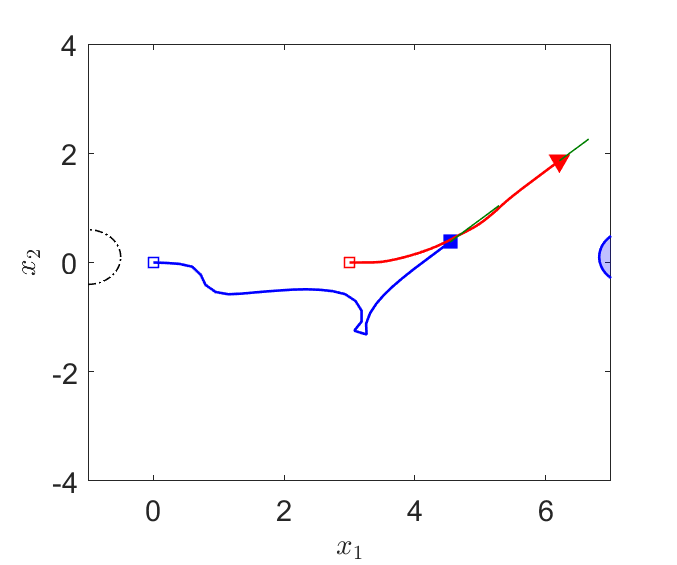}}
    \caption{Simulation results for the instance where the pursuer initially moves towards the local minimum that is dominated by the RCS cost-map before capturing the evader ($Q=1$).}
    \label{fig:Desensi_NoDecept_3}
    \vspace*{-5pt}
\end{figure*}

\begin{figure*}[htb!]
    \centering
    \subfigure[$t = 1.2$]{\includegraphics[width = 0.32\textwidth]{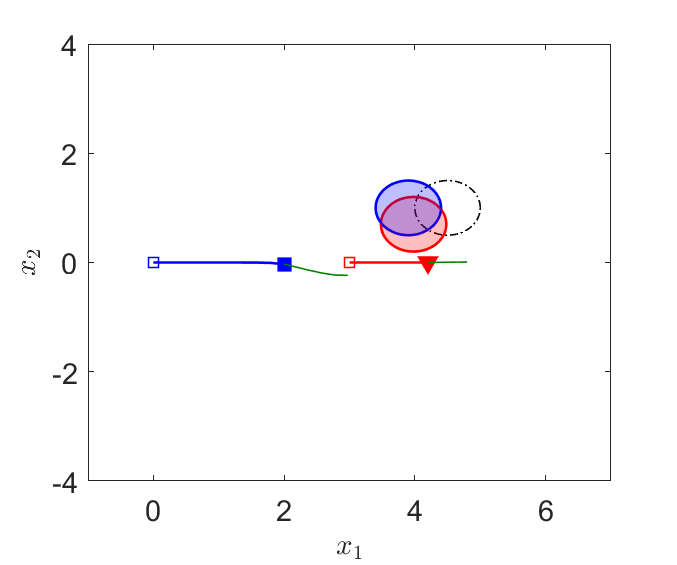}}
    \subfigure[$t = 2.3$]{\includegraphics[width = 0.32\textwidth]{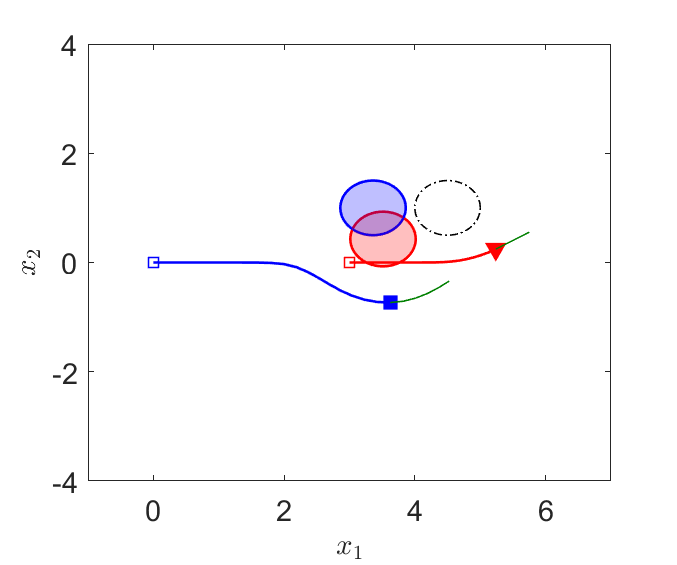}}
    \subfigure[$t = 10.0$]{\includegraphics[width = 0.32\textwidth]{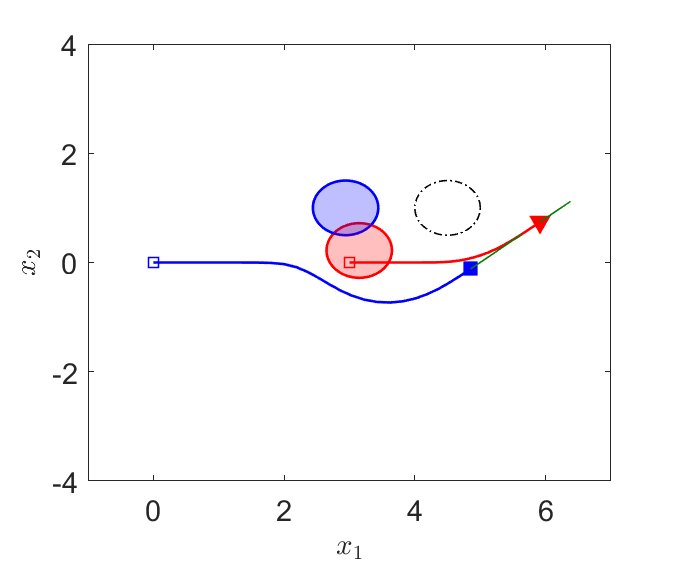}}
    \caption{Simulation results for an instance where the pursuer penalizes the proposed risk estimate with $Q=2.5$, and successfully captures the evader while avoiding the obstacle. The uncertain parameter is the obstacle's heading.}
    \label{fig:Desensi_heading}
    \vspace*{-5pt}
\end{figure*}

\subsection{Desensitization Against a Non-Deceptive Evader}
\label{subsec:sim_nodecept}

The effectiveness of desensitization is initially analyzed against a non-deceptive evader.
Initially, the case where the pursuer does not desensitize, which can be captured by setting $Q=0$, 
is examined in Fig.~\ref{fig:DOC_colli}.
The simulation parameters are $\epsilon = 0.3$, $r_o = 0.75$, $N = 10$, $\Delta t = 0.1$.
The initial positions are $x_p^0 = (0,0)$, $x_e^0 = (3,0)$, and $x_w^0 = (2,1.15)$.
The speeds are set to $u_c = 1$, $v_c = 0.6$, $\bar{\rho} = [0~-0.25]$, and ${\rho} = [0~-0.35]$.
It is assumed that $\rho_2$ is the only uncertain parameter.
In this case, the pursuer ends up colliding with the obstacle, and the result is presented in Fig. \ref{fig:DOC_colli}. 
The blue line denotes the trajectory traced by the pursuer with the markers denoting the initial and the instantaneous position of the pursuer.
Similarly, red color corresponds to the evader.
The blue circle is the instantaneous nominal position of the obstacle, as estimated by the pursuer, and the red circle is that of the actual obstacle, as estimated by the evader.
The black dashed circle indicates the initial position of the obstacle.
The green curves denote the instantaneous $N$-step look ahead trajectories for both players based on the RHC control law.
As can be seen in Fig. \ref{fig:DOC_colli}, the lack of information regarding the whereabouts of the true obstacle and the reliance on its nominal model led to a collision in this case.

Figure~\ref{fig:Desensi_NoDecept_1} presents the results obtained when the pursuer employs a desensitized strategy with $Q=1$ for the case presented in Fig.~\ref{fig:DOC_colli}. 

The fact that the pursuer minimizes a risk estimate at every instant can be inferred from the instantaneous look-ahead trajectory (the green curves) in Fig.~\ref{fig:Desensi_NoDecept_1}(a), which takes the pursuer away from the obstacle, as opposed to taking it along the line-of-sight.
From Fig.~\ref{fig:Desensi_NoDecept_1}(b), it can be observed that the pursuer navigates around the obstacle by maintaining a safe distance before the pursuer turns its full attention towards capturing the evader in pure pursuit, which is shown in Fig.~\ref{fig:Desensi_NoDecept_1}(c).

Figure~\ref{fig:Desensi_NoDecept_2} showcases the results for an instance where the obstacle is moving along the $x_1$-axis.
The related simulation parameters that are different from the previous case are $N = 5$, $\Delta t = 0.2$, $x_w^0 = (4,0.1)$, $\bar{\rho} = [-0.25~0]$, ${\rho} = [-0.35~0]$; it is assumed that $\rho_1$ is the only uncertain parameter.
From Fig.~\ref{fig:Desensi_NoDecept_2}(a), it can be observed that the evader, knowing the true position of the obstacle, maneuvers around it without colliding.
Figures~\ref{fig:Desensi_NoDecept_2}(a) and \ref{fig:Desensi_NoDecept_2}(b) indicate that the pursuer takes a risk-averse trajectory by executing an early turn maneuver to avoid collision with the true obstacle. 
Similar to what is observed in the earlier case, after safely avoiding the true obstacle, the pursuer engages in a pure pursuit with the evader, as can be seen in Fig. \ref{fig:Desensi_NoDecept_2}(c).
This case could be considered for future research involving the estimation of unknown model parameters using the trajectories of the more informed player.
Since the pursuer knows that the evader is aware of the obstacle's true model, and can observe the evader's trajectory, an inference problem can be formulated to predict the true model, which is beyond the scope of this paper.

Another case of desensitization against a non-deceptive evader is analyzed in Fig.~\ref{fig:Desensi_NoDecept_3}.
The related simulation parameters that are different from the case discussed in Fig.~\ref{fig:Desensi_NoDecept_2} are $x_w^0 = (-1,0.1)$, $\bar{\rho} = [1.3~0]$, ${\rho} = [1.4~0]$; it is assumed that $\rho_1$ is the only uncertain parameter.
Note that the obstacle (nominal/true) is faster compared to both players.
The pursuer successfully avoids colliding with the obstacle with the RCS regularizer driving the pursuer away from the obstacle's center line along the $x_1$-axis.
Once the pursuer avoids collision, Fig.~\ref{fig:Desensi_NoDecept_3}(a) suggests that the pursuer moves towards the local minimum that is along the axis parallel to $x_2$-axis and below the nominal obstacle (blue circle).
This local minimum exists due to the fact that there is no uncertainty in $\rho_2$, resulting in a valley in the RCS regularizer cost map.
Since the nominal obstacle is faster compared to the pursuer, as the obstacle moves past the pursuer, the poles in the RCS field (see Fig.~\ref{fig:RCS_norm}(b)), that are parallel to the $x_1$-axis, force the pursuer to move away from the nominal obstacle (see Fig.~\ref{fig:Desensi_NoDecept_3}(b)).
As shown in Fig.~\ref{fig:Desensi_NoDecept_3}(c), when the nominal obstacle moves past the pursuer, and the strength of the RCS field is reduced, the pursuer then employs pure pursuit to capture the evader.

In a series of simulations demonstrating the effectiveness of desensitization against a non-deceptive evader, we finally consider the case where the obstacle's heading ($\psi_\rho$) is the only uncertain parameter 
(see Fig.~\ref{fig:Desensi_heading}).
The related simulation parameters that are different from the case discussed in Fig. \ref{fig:Desensi_NoDecept_1} are $x_w^0 = (4.5,1)$, $\|\rho\|_2 = 0.3$. 
The nominal value of the obstacle's heading is $180^\circ$, and its true value is $-150^\circ$.
Note that the obstacle's total speed ($\|\rho\|_2$) is not an uncertain parameter.
As discussed in Section \ref{subsec:rcs_fields}, since the RCS field in this case is orthogonal to the obstacle's nominal velocity vector (which is along the negative $x_1$-axis), the pursuer maintains a safe distance in the direction parallel to the $x_2$-axis to effectively avoid the true obstacle, and successfully capture the evader.

\subsection{Desensitization Against a Deceptive Evader}

In this subsection, we examine the performance of the desensitized strategy against a deceptive evader.
For the remaining set of simulations in this subsection, the evader employs pure deception, i.e., $\alpha_o = 0$ and $\alpha_d = 1$, as described in ($\mathcal{E}_d$).
Note that the evader considers a pure pursuit feedback strategy for the pursuer to construct  her deceptive strategy. 

\begin{figure*}[htb!]
    \centering
    \subfigure[$t = 0.6$]{\includegraphics[width = 0.32\textwidth]{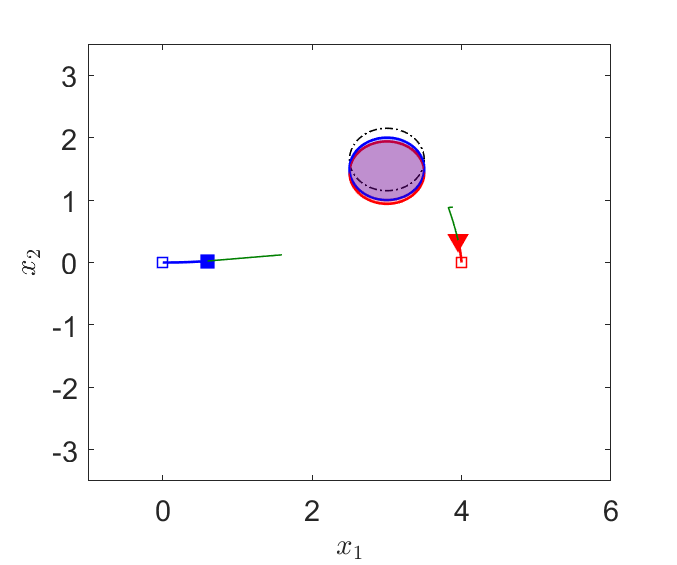}}
    \subfigure[$t = 1.7$]{\includegraphics[width = 0.32\textwidth]{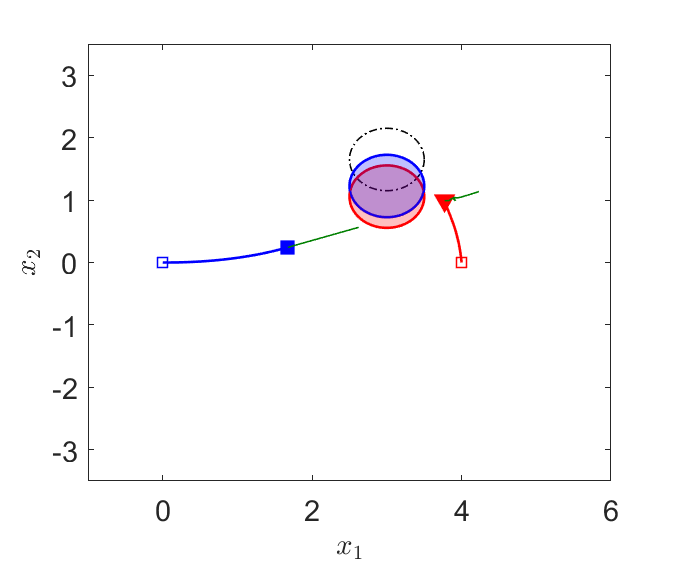}}
    \subfigure[$t = 2.6$]{\includegraphics[width = 0.32\textwidth]{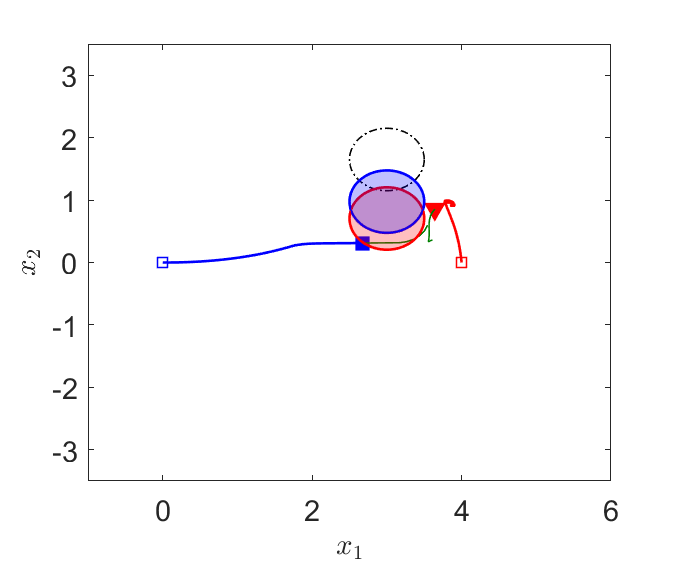}}
    \caption{An instance where the pursuer solves $\mathcal{P}_o$ against a deceptive evader, but ends up colliding with the obstacle ($Q=0$).}
    \label{fig:Decept_colli}
    \vspace*{-5pt}
\end{figure*}

\begin{figure*}[htb]
    \centering
    \subfigure[$t = 1.3$]{\includegraphics[width = 0.32\textwidth]{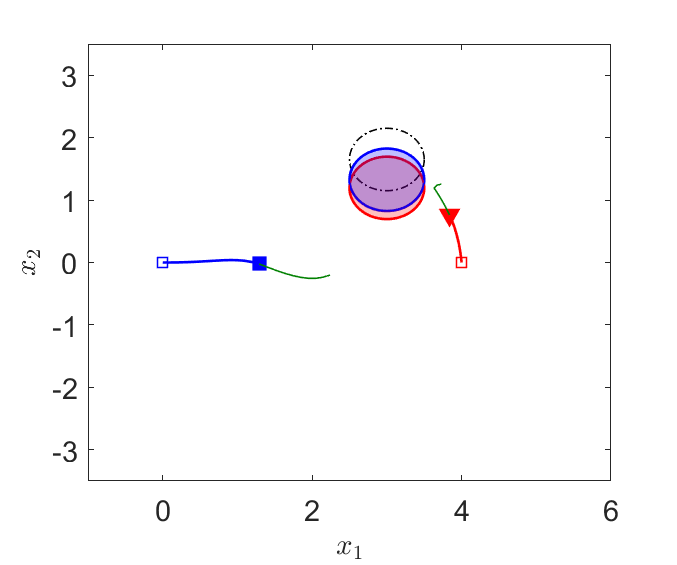}}
    \subfigure[$t = 3.8$]{\includegraphics[width = 0.32\textwidth]{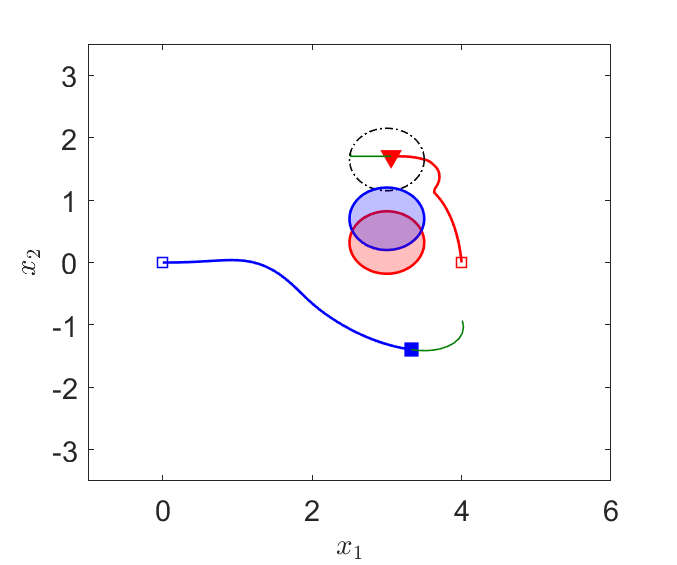}}
    \subfigure[$t = 8.4$]{\includegraphics[width = 0.32\textwidth]{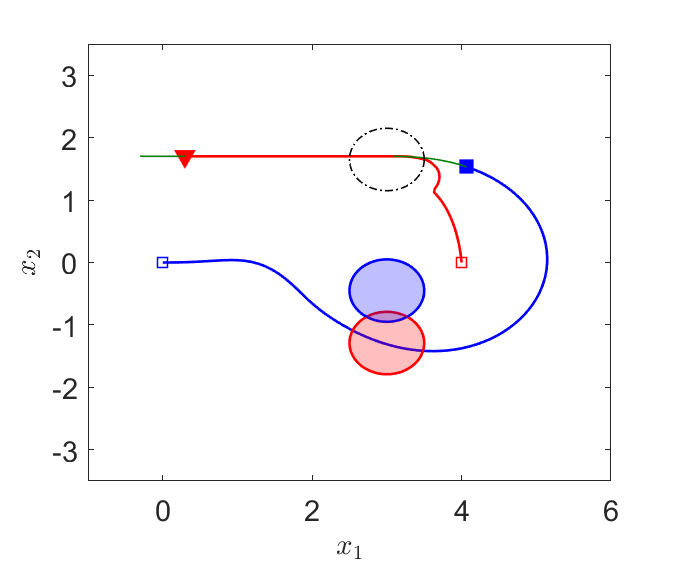}}
    \caption{Simulation results for the instance where the pursuer employs the proposed desensitized strategy to successfully capture a deceptive evader ($Q=0.5$).}
    \label{fig:Desensi_Decept}
    \vspace*{-5pt}
\end{figure*}

\begin{figure*}[htb]
    \centering
    \subfigure[$t = 1.2$]{\includegraphics[width = 0.32\textwidth]{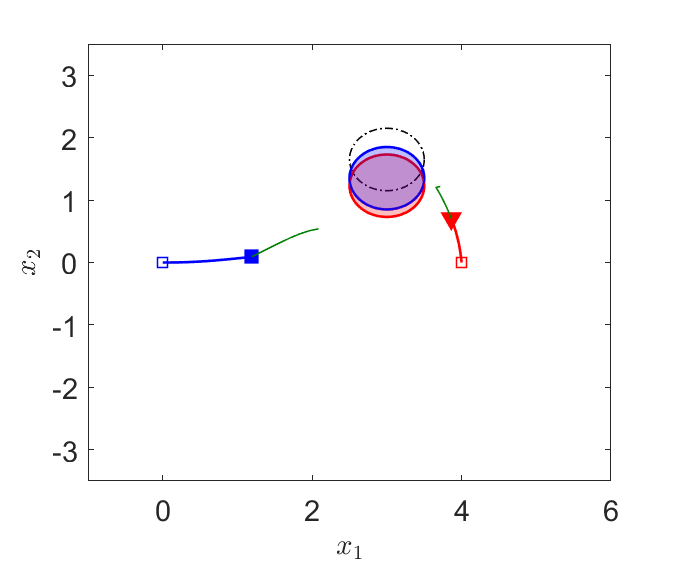}}
    \subfigure[$t = 2.3$]{\includegraphics[width = 0.32\textwidth]{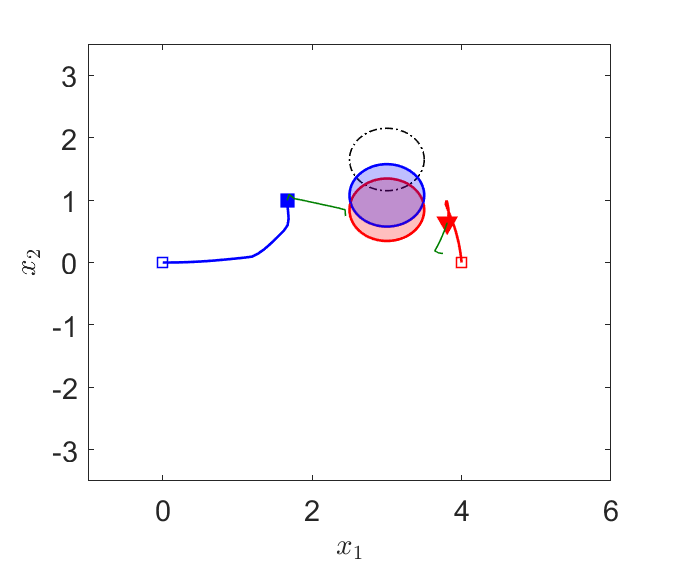}}
    \subfigure[$t = 10.0$]{\includegraphics[width = 0.32\textwidth]{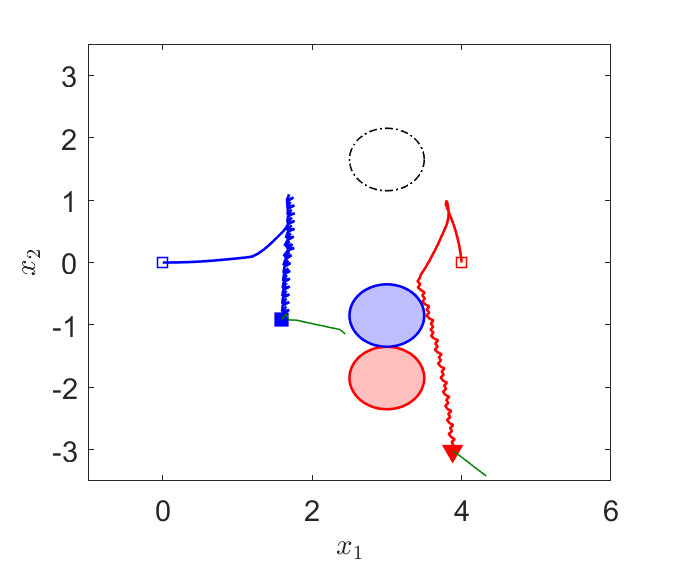}}
    \caption{Simulation results for the instance where the pursuer employs the proposed desensitized strategy but gets stuck in a local minimum ($Q=0.5$).}
    \label{fig:Desensi_Decept_LM}
    \vspace*{-5pt}
\end{figure*}

The simulation parameters that are common for the results, shown in Figs. \ref{fig:Decept_colli}-\ref{fig:Desensi_Decept_LM}, include $N = 10$, $\Delta t = 0.1$, $x_p^0 = (0,0)$, $x_e^0 = (4,0)$, and $x_w^0 = (3,1.65)$, $u_c = 1$, $v_c = 0.6$, $\bar{\rho} = [0~-0.25]$, ${\rho} = [0~-0.35]$.
Figure~\ref{fig:Decept_colli} presents the case where the pursuer collides with the true obstacle as the evader lures the pursuer, which solves the nominal game ($\mathcal{P}_o$), with a deceptive strategy.
Therefore, as can be observed in Fig.~\ref{fig:Decept_colli}, the evader initially moves towards the true obstacle, and then maintains a position that would put the true obstacle between both players.
As a result, the pursuer ends up hitting the true obstacle.

Figure~\ref{fig:Desensi_Decept} showcases the player's trajectories for the instance where the pursuer employs the proposed desensitized strategy with $Q = 0.5$. 
For this case, it is assumed that both $\rho_1$ and $\rho_2$ are the uncertain parameters against which the pursuer is constructing the proposed risk estimate.
As the pursuer maneuvers around the true/nominal obstacle, from Fig.~\ref{fig:Desensi_Decept}(b), it can be noted that the evader constantly shifts to place the true obstacle along the line-of-sight.
However, the pursuer follows a risk-averse trajectory, similar to the one observed in the case of a non-deceptive evader in Fig.~\ref{fig:Desensi_NoDecept_1}, and counteracts the evader's deception strategy.

Finally, we analyze a case that demonstrates the implementation challenges observed with the proposed approach to construct risk-sensitive strategies. 
From Sections \ref{subsec:rcs_fields} and \ref{subsec:sim_nodecept}, it can be inferred that the risk-sensitive trajectories are primarily driven by the associated RCS fields, which are dependent on the type of uncertainty assumed. 
In this regard, consider the case where the pursuer employs a desensitized strategy against a deceptive evader while assuming uncertainty in $\rho_2$ only.
The results are shown in Fig.~\ref{fig:Desensi_Decept_LM} (simulation parameters for Figs. \ref{fig:Desensi_NoDecept_1} and \ref{fig:Desensi_Decept_LM} are the same).
From Figs.~\ref{fig:Desensi_Decept_LM}(a) and \ref{fig:Desensi_Decept_LM}(b), it can be observed that the pursuer is driven towards a local minimum.
The RCS field (a $90^\circ$ transformation of the field shown to Fig.~\ref{fig:RCS_norm}(b)) is such that the pursuer is stuck at the local minimum while the evader continuously attempts to drive the pursuer towards the true obstacle, as can be seen in Fig. \ref{fig:Desensi_Decept_LM}(c).
The local minimum in Fig.~\ref{fig:Desensi_Decept_LM} may be addressed by tuning: 1) the uncertain parameter set that the pursuer considers for constructing the risk estimate; 2) the chosen simulation specifications for receding horizon control, including $N$ and $\Delta t$; and 3) the weight on the RCS norm $Q$.

For the more informed player (Evader), it is important to understand the \emph{relevance of deception}.
From the results presented in Figs. \ref{fig:Decept_colli}-\ref{fig:Desensi_Decept_LM}, it can be inferred that the evader has the incentive to employ a deceptive strategy to trick the pursuer. 
However, it is obvious that when the obstacle starts from a faraway point, it is ineffective for the evader to resort to reducing the distance between the pursuer and the true obstacle, which could lead to an early capture.
In such a case, the evader needs to constantly evaluate the incentive to employ a deceptive strategy, which then becomes dependent on the chosen pursuer's feedback strategy.
Subsequently, the relevance of deception in the pursuer-evader-obstacle problem discussed in this paper can be addressed as follows.
Given the instantaneous positions of the players and that of the obstacle, along with the chosen feedback strategy of the pursuer $\hat{u}$, deception is relevant for the evader if there exists at least one evader trajectory from her reachable set that drives the pursuer towards a collision with the obstacle before the evader is captured.
This opens the opportunity for feedback control-based reachable sets that could address safety in multi-agent problems, a topic which, however, is beyond the scope of this paper.


\section{Conclusion}
\label{sec:conclude}

This paper investigates the aspects of risk aversion and deception in differential games with asymmetric information.
Using the example of pursuit-evasion in an environment with an uncertain obstacle, it has been shown that information advantage provides an incentive for the more informed player to employ deception against a less informed player.
In order to address planning under lack of information and deceptive strategies employed by the adversary, desensitized strategies that minimize the risk of constraint violation are proposed for the player with information disadvantage.
The simulations suggest that desensitized strategies can mitigate both information uncertainty and deception in strategic engagements involving a more informed adversary.


\bibliographystyle{plain}
\bibliography{references}

\end{document}